\documentclass[12pt]{article}      
 \usepackage[top=1in, bottom=1in, left=1in, right=1in]{geometry}
\usepackage{graphicx}
\usepackage{authblk}
\usepackage{natbib}
 \usepackage{pdflscape}
\usepackage{lscape}
\usepackage{amssymb}\usepackage{amsmath,amsthm}
\usepackage{amsmath} \usepackage{multirow}

\numberwithin{equation}{section} 
\usepackage{amscd,epstopdf}
\usepackage{graphicx}\usepackage{multirow}
\usepackage{float}
\usepackage{subcaption}

\usepackage{afterpage} \usepackage{rotating}
\usepackage{enumerate}
 
 \usepackage{setspace}
\usepackage{indentfirst} 
\usepackage{graphicx}
\usepackage{bibentry}

\begin{document}

\title{The E-Bayesian Estimation and its E-MSE of Lomax distribution
	under different loss functions}


\author[1]{Kaiwei Liu  }
\author[1]{Yuxuan Zhang }

\affil[1]{{\footnotesize Department of Mathematics, Beijing Jiaotong University, Beijing 100044, China}}
   
 \renewcommand\Authands{ and }
 \date{}
 \maketitle

\begin{abstract}
This paper studies the E-Bayesian (expectation of the Bayesian estimation) estimation of the parameter of Lomax distribution based on different loss functions. Under different loss functions, we calculate the Bayesian estimation of the parameter and then calculate the expectation of the estimated value to get the E-Bayesian estimation. To measure the estimated error, the E-MSE (expected mean squared error) is introduced. And the formulas of E-Bayesian estimation and E-MSE are given. By applying Markov Chain Monte Carlo technology, we analyze the performances of the proposed methods. Results are compared on the basis of E-MSE. Then, cases of samples in real data sets are presented for illustration. In order to test whether the Lomax distribution can be used in analyzing the datasets, Kolmogorov– Smirnov tests are conducted. Using real data, we can get the maximum likelihood estimation at the same time and compare it with E-Bayesian estimation. At last, we get the results of the comparison between Bayesian and E-Bayesian estimation methods under three different loss functions.

\end{abstract}


\section{Introduction}

Lomax distribution is a widely used life distribution in reliability and life test research, especially in the analysis of life test data processing in medicine, biological sciences, and engineering sciences. This distribution has monotonically increasing and decreasing failure rates. 

The probability density function (PDF) of Lomax distribution is as follows:
 \begin{eqnarray}\label{pdf}
f(x;\alpha,\lambda)=\frac{\alpha}{\lambda}(1+\frac{x}{\lambda})^{-(\alpha+1)},\quad x>0.
\end{eqnarray}

Hence, the cumulative distribution function (CDF) is given by

 \begin{eqnarray}\label{cdf}
 F(x;\alpha,\lambda)=1-(1+\frac{x}{\lambda})^{-\alpha },\quad x>0; \quad \alpha,\lambda>0,
 \end{eqnarray}
where $\alpha$ is the shape parameter and $\lambda$ is the scale parameter. The reliability function $R(t)$ and the hazard rate function $h(t)$ for Lomax distribution are, respectively, given by
\begin{eqnarray}\label{failure rate function}
R(t)=(1+\frac{t}{\lambda})^{-\alpha }, \quad\alpha,\lambda>0, \quad t>0
\end{eqnarray}
and
\begin{eqnarray}
 h(t)=\frac{\alpha}{\lambda}(1+\frac{t}{\lambda})^{-1 },\quad \alpha,\lambda>0,\quad t>0.
\end{eqnarray}

When $X\sim Lomax(\alpha,\lambda)$, the expection of $X$ is
\begin{eqnarray}
E(X)=\frac{\lambda}{\alpha-1}, \quad\alpha>1
\end{eqnarray}
and the variance of $X$ is
\begin{eqnarray}
Var(X)=\frac{\alpha{\lambda}^2}{{(\alpha-1)}^2{(\alpha-2)}},\quad\alpha>2.
\end{eqnarray}

The line charts of the probability density function of Lomax distribution are shown in Figure 1 and Figure 2.
\begin{figure}[h]
	\centering
	\includegraphics[width=10cm,height=10cm]{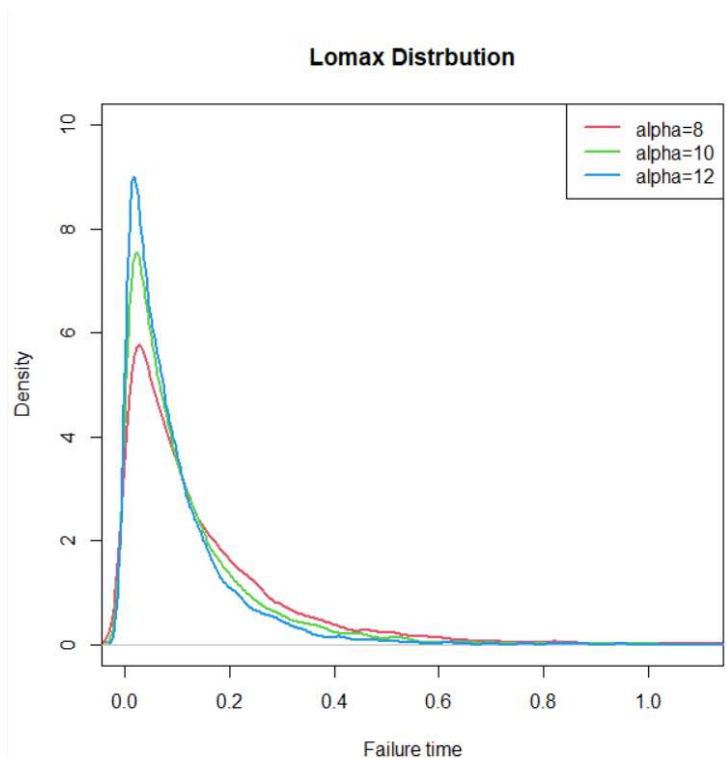}
	\caption{PDF of Lomax distribution with different values of $\alpha$}
\end{figure}

\begin{figure}[h]
	\centering
	\includegraphics[width=10cm,height=10cm]{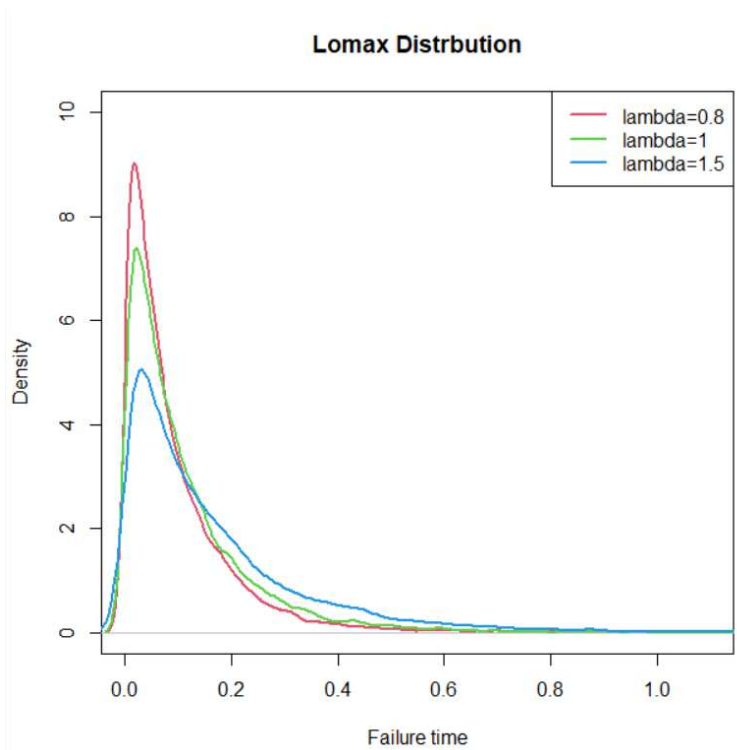}
	\caption{PDF of Lomax distribution with different values of $\lambda$}
\end{figure}

When it comes to Figure 1 and Figure 2, the plots of the probability density functions, we can observe the characteristics of the line charts under the different values of $\alpha$ and $\lambda$. We get the charts when $\alpha=8,10,12$ as $\lambda=1$ and when $\lambda=0.8,1,1.5$ as $\alpha=10$. The probability density is concentrated in places where the value of time is smaller as the parameter $\alpha$ increases. And as the parameter $\lambda$ increases, probability density decreases in places where the value of time is smaller.

If $x=(x_1, x_2, \cdots, x_n)$ is a set of samples obeying the Lomax distribution, when the shape parameter is unknown and the scale parameter is known, the likelihood function is as follows:
\begin{eqnarray}
L(x|\alpha)=\displaystyle\prod_{i=1}^n [\frac{\alpha}{\lambda}(1+\frac{x_i}{\lambda})^{-(\alpha+1)}]\propto{\alpha}^n exp{(-\alpha T),}
\end{eqnarray}
where $T=\displaystyle\sum_{i=1}^n \ln(1+\frac{x_i}{\lambda})$.

The parameters of the Lomax distribution can be estimated in different ways under different conditions. \cite{Labban2019} discussed the two-parameter estimation of Lomax distribution. \cite{AlSobhi2013} gave the estimation of the parameters of Lomax distribution under general progressive censoring.

At the same time, composite distribution functions of Lomax distribution have been studied too. \cite{Cordeiro2015} and \cite{Alzahrani2014} gave the research of the Gamma-Lomax distribution and the Poisson-Lomax distribution.  \cite{Almetwally2020} gave the parameter estimation and stress-strength model of power Lomax distribution. 

Different estimation methods are used in this paper. Each method has its own disadvantages, such as computational complexity, estimation error and so on. Therefore, a new method, E-Bayesian method, is used in this study.

Bayesian statistical inference method depends on the choice of prior distribution and loss function. But the prior distribution parameter may depend on the hyperparameters. In this case, we usually use the hierarchical Bayesian method. \cite{Smith1972} first proposed the idea of the hierarchical prior distribution. The hierarchical Bayesian method needs two stages to complete the setting of the prior distribution, so it is more robust than the Bayesian method. \cite{Han1997} proposed a method to construct hierarchical prior distribution. Recently, the hierarchical Bayesian method has been applied to data analysis. For more details, refer to \cite{Arnold2010}. However, hierarchical Bayesian estimation often involves the calculation of complex integrals. In practical application, it is difficult to use the hierarchical Bayesian method to calculate integrals. Some calculation methods such as the Markov chain Monte Carlo (MCMC) method are available especially the Metropolis-Hastings (MH) algorithm. For more details we can refer to \cite{Hastings1970} and \cite{Cowles1995}.

On the other hand, loss function plays an important role in Bayesian methods. In Bayesian reasoning, square error loss is most commonly used. This kind of loss function is symmetrical and gives equal weight to both overestimation and underestimation. It is well known that in many cases, it may not be appropriate to use symmetric loss function, especially when positive error and negative error have different consequences. One of the most commonly used asymmetric loss functions is the LINEX (linear exponential) loss function. \cite{Varian1975} introduced this point, and \cite{Zellner1986} studied the further properties of this loss function. In many practical cases, it seems more realistic to express loss by ratio. In this case, \cite{Parsian1996} pointed out that another useful asymmetric loss function is the entropy loss function. For parameter estimation under entropy loss function, see \cite{Parsian1996}. Research on Bayesian estimation under different loss functions is given in \cite{Han2018}.

The E-Bayesian estimation of parameters of Lomax distribution based on different loss functions is studied in this article.

\subsection{Maximum likelihood estimation of Lomax distribution}
Maximum likelihood estimation (MLE) is an important and universal method to get the estimation. The maximum likelihood method explicitly uses the probability model, and its goal is to find the phylogenetic tree that can produce observation data with high probability. The maximum likelihood method is the representative of a class of completely statistical based phylogenetic tree reconstruction methods. In this section, we will calculate the MLE of Lomax distribution as the contrast of Bayesian estimation and E-Bayesian estimation.

Refer to \cite{Han2019a}, the maximum likelihood estimate for $\alpha$ is derived as follows:

From (1.7) we conclude
\begin{eqnarray}
L(x|\alpha)=(\frac{\alpha}{\lambda})^n\displaystyle\prod_{i=1}^n [(1+\frac{x_i}{\lambda})^{-(\alpha+1)}]
\end{eqnarray}
where $T=\displaystyle\sum_{i=1}^n \ln(1+\frac{x_i}{\lambda})$.

Logarithm the two sides of the equation, we get
\begin{eqnarray}
\ln L(x|\alpha)=n\ln\alpha-n\ln\lambda-(\alpha+1)T.
\end{eqnarray}

Take the partial derivatives of $\alpha$ on both sides of the equation,
\begin{eqnarray}
\frac{\partial\ln L(x|\alpha)}{\partial\alpha}=\frac{n}{\alpha}-T.
\end{eqnarray}

Let 
\begin{eqnarray}
\frac{\partial\ln L(x|\alpha)}{\partial\alpha}=0
\end{eqnarray}
we can compute the MLE of $\alpha$ as
\begin{eqnarray}
\hat{\alpha}_{ML}=\frac{n}{T},
\end{eqnarray}
which is used as a comparison of Bayesian estimation.

\section{Bayesian Estimation of $\alpha$} 
In this section, we will calculate the Bayesian estimation of alpha under three loss functions. According to \cite{Ali2013}, let $x=(x_1, x_2, \cdots, x_n)$ be a set of samples obeying the Lomax distribution, for any prior distribution of parameter $\alpha$, have the following conclusions:

\begin{enumerate}[1)]
	\item Under the square error loss (SEL) function, $L_1 (\alpha,\delta)=(\alpha-\delta)^2$, the Bayesian estimation of $\alpha$ is $\hat{\alpha}_{B1}(x)=E(\alpha|x)$.
	\item Under the K-loss (KL) function, $L_2 (\alpha,\delta)=(\sqrt{\alpha/\delta}-\sqrt{\delta/\alpha})^2$, the Bayesian estimation of $\alpha$ is $\hat{\alpha}_{B2}(x)=\sqrt{{E(\alpha|x)}/{E(\alpha^{-1}|x)}}$.
	\item Under the entropy loss (EL) function, $L_3 (\alpha,\delta)=\delta/\alpha-\ln(\delta/\alpha)-1$, the Bayesian estimation of $\alpha$ is $\hat{\alpha}_{B3}(x)=[E(\alpha^{-1}|x)]^{-1}$.
\end{enumerate}

In this section, we consider that the Bayesian estimation of the unknown parameter $\alpha$ is based on the above loss functions. The Bayesian approach requires the specification of the prior PDF of the parameter $\alpha$. Because $\alpha>0,\lambda>0$, it is assumed that the PDF of Gamma distribution is the prior PDF as follows:

\begin{eqnarray}
\pi(\alpha)=\frac{b^a}{\Gamma(a)} \alpha^{a-1} e^{-b\alpha}.
\end{eqnarray}
 
When $\lambda$ is fixed, and $X=(X_{1:n}, X_{2:n}, \cdots, X_{n:n})$, $x=(x_{1:n}, x_{2:n}, \cdots, x_{n:n})$, from (2.2) and (2.3), the posterior PDF of $\alpha$ is given as follows:

\begin{eqnarray}
\pi(\alpha|x)=\frac{L(\alpha)\pi(\alpha)}{\int_0^{+\infty}L(\alpha)\pi(\alpha)d\alpha}.
\end{eqnarray}\\

So we can get the Bayesian estimation of $\alpha$ under three loss functions.

\subsection{Bayesian Estimation of $\alpha$ under SEL function}
Under SEL function, $L_1 (\alpha,\delta)=(\alpha-\delta)^2$, the Bayesian estimation of $\alpha$ is
\begin{eqnarray}
\begin{aligned}
\hat{\alpha}_{B1}&=E[\alpha|x]=\int_{0}^{+\infty}\alpha\pi(\alpha|x)d\alpha\\&
=\frac{\displaystyle\int_{0}^{+\infty}\alpha L(\alpha)\pi(\alpha)d\alpha}{\displaystyle\int_0^{+\infty}L(\alpha)\pi(\alpha)d\alpha}
=\frac{\displaystyle\int_{0}^{+\infty}\alpha^{a+n} e^{-(b+T)\alpha}d\alpha}{\displaystyle\int_{0}^{+\infty}\alpha^{a+n-1} e^{-(b+T)\alpha}d\alpha}\\&
=\frac{\displaystyle\frac{\Gamma(a+n+1)}{(b+T)^{a+n+1}}}{\displaystyle\frac{\Gamma(a+n)}{(b+T)^{a+n}}}
=\frac{\Gamma(a+n)(a+n)}{\Gamma(a+n)(b+T)}=\frac{a+n}{b+T}.
\end{aligned}
\end{eqnarray}

\subsection{Bayesian Estimation of $\alpha$ under KL function}
Under KL function, the Bayesian estimation of $\alpha$ is
\begin{eqnarray}
\begin{aligned}
\hat{\alpha}_{B1}&=\sqrt{{E(\alpha|x)}/{E(\alpha^{-1}|x)}}=\sqrt{\frac{\displaystyle\int_{0}^{+\infty}\alpha\pi(\alpha|x)d\alpha}{\displaystyle\int_{0}^{+\infty}\alpha^{-1}\pi(\alpha|x)d\alpha}}\\&
=\sqrt{\frac{\displaystyle\int_{0}^{+\infty}\alpha L(\alpha)\pi(\alpha)d\alpha}{\displaystyle\int_0^{+\infty}L(\alpha)\pi(\alpha)d\alpha}\frac{\displaystyle\int_{0}^{+\infty} L(\alpha)\pi(\alpha)d\alpha}{\displaystyle\int_0^{+\infty}\alpha^{-1}L(\alpha)\pi(\alpha)d\alpha}}
\\&
=\sqrt{\frac{\Gamma(a+n)(a+n)}{\Gamma(a+n)(b+T)}\frac{\Gamma(a+n-1)(a+n-1)}{\Gamma(a+n-1)(b+T)}}\\&
=\frac{\sqrt{(a+n)(a+n-1)}}{b+T}.
\end{aligned}
\end{eqnarray}

\subsection{Bayesian Estimation of $\alpha$ under EL function}
Under EL function, the Bayesian estimation of $\alpha$ is
\begin{eqnarray}
\begin{aligned}
\hat{\alpha}_{B1}&=[E(\alpha^{-1}|x)]^{-1}=\displaystyle(\int_{0}^{+\infty}\alpha^{-1}\pi(\alpha|x)d\alpha)^{-1}\\&
=\displaystyle(\frac{\displaystyle\int_{0}^{+\infty}\alpha^{-1} L(\alpha)\pi(\alpha)d\alpha}{\displaystyle\int_0^{+\infty}L(\alpha)\pi(\alpha)d\alpha})^{-1}
=\frac{\displaystyle\int_{0}^{+\infty}\alpha^{a+n-1} e^{-(b+T)\alpha}d\alpha}{\displaystyle\int_{0}^{+\infty}\alpha^{a+n-2} e^{-(b+T)\alpha}d\alpha}\\&
=\frac{\Gamma(a+n-1)(a+n-1)}{\Gamma(a+n-1)(b+T)}=\frac{a+n-1}{b+T},
\end{aligned}
\end{eqnarray}
where $T=\displaystyle\sum_{i=1}^n \ln(1+\frac{x_i}{\lambda})$.

 \section{ E-Bayesian Estimation of $\alpha$}
According to \cite{Han1997}, the prior parameters $a$ and $b$ should be selected to guarantee
that $\pi(\alpha)$ is a decreasing function of $\alpha$. The derivative of $\pi(\alpha)$ with respect to $\alpha$ is
given by
\begin{eqnarray}
\frac{d\pi(\alpha)}{d\alpha}=\frac{b^a}{\Gamma(a)}\alpha^{a-2}e^{-b\alpha}\{a-1-b\alpha\},\quad 0<a<1,b>0,
\end{eqnarray}
where $\Gamma(a)=\int_{0}^{+\infty}x^{a-1}e^{-x}dx$ is the gamma function, and hyper parameters $a>0$ and $b>0$.

The prior PDF $\pi(\alpha)$ is a decreasing function of $\alpha$. We can assume that the hyperparameters called $a$ and $b$ are independent with bi-variate PDF given by
\begin{eqnarray}
\pi(a,b)=\pi_1(a)\pi_2(b).
\end{eqnarray}

We define that $a>0, b>0$, and $\alpha>0$, it follows $0<a<1, b>0$ due to $\frac{d\pi(\alpha)}{d\alpha}< 0$, and
therefore $\pi(\alpha|a, b)$ is a decreasing function of $\lambda$. Given $0<a<1$, the larger $b$ is, the thinner
the tail of the Gamma density function will be. Considering the robustness of Bayesian estimate in \cite{Berger1985}, the thinner tailed prior distribution often reduces the robustness of Bayesian
estimate. Accordingly, $b$ should not be larger then a given upper bound $c$, where $c>0$ is
a constant to be determined. Thereby, the hyper parameters $a$ and $b$ should be selected
with the restriction of $0<a<1$ and $0<b<c$. How to determine the constant $c$ would be
described later in application example.

The E-Bayesian estimation of the parameter $\alpha$ is given by
\begin{eqnarray}
\hat{\alpha}_{EB}=E[\hat{\alpha}_{B}|x]=\iint_{Q}\hat{\alpha}_B(a,b)\pi(a,b) \, da\,db,
\end{eqnarray}
in which $Q$ is the set of all possible values of the parameters $a$ and $b$. $\hat{\alpha}_B(a,b)$ is the Bayesian estimations of the parameter $\alpha$. We indicates that $\hat{\alpha}_{EB}=E[\hat{\alpha}_{B}(a,b)]$.

If $X_{1}, X_{2}, \cdots , X_{n}$ are the failure sample observations from the Lomax distribution (1.1), the likelihood function is given by (1.7), the prior density function $\pi(\alpha |a,b)$ of $\alpha$ is given by (2.2). Refer to \cite{Han2020}, the prior density function $\pi(a, b)$ of $a$ and $b$ is given by
\begin{eqnarray}
\pi(a,b)=\frac{1}{c},\quad 0<a<1, 0<b<c.
\end{eqnarray}

The definition of E-Bayesian estimation was originally addressed by \cite{Han2007}.

\subsection{E-Bayesian Estimation of $\alpha$ under SEL function}
The E-Bayesian estimate of $\alpha$ under SEL function can be obtained from (2.3), (3.3), and (3.4) as follows:
\begin{eqnarray}
\begin{aligned}
\hat{\alpha}_{EB1}&=\iint_{Q}\hat{\alpha}_{B1}(a,b)\pi(a,b) \, da\,db
\\&=\iint_{Q}\frac{a+n}{b+T}\frac{1}{c} \, da\,db=\frac{1}{c}\int_{0}^{1}(a+n)da\int_{0}^{c}\frac{1}{b+T}db\\&=\frac{2n+1}{2c}\ln(\frac{T+c}{T}),
\end{aligned}
\end{eqnarray}
from which we can get the E-Bayesian estimate of $\alpha$ is
\begin{eqnarray}
\hat{\alpha}_{EB1}=\frac{2n+1}{2c}\ln(\frac{T+c}{T}).
\end{eqnarray}

\subsection{E-Bayesian Estimation of $\alpha$ under KL function}
The E-Bayesian estimate of $\alpha$ under SEL function can be obtained from (2.4), (3.3), and (3.4) as follows:
\begin{eqnarray}
\begin{aligned}
\hat{\alpha}_{EB2}&=\iint_{Q}\hat{\alpha}_{B2}(a,b)\pi(a,b) \, da\,db\\&
=\iint_{Q}\frac{\sqrt{(a+n)(a+n-1)}}{b+T}\frac{1}{c} \, da\,db\\&=\frac{1}{c}\int_{0}^{1}\sqrt{(a+n)(a+n-1)}da\int_{0}^{c}\frac{1}{b+T}db\\&=\frac{1}{c}\ln(\frac{T+c}{T})\int_{0}^{1}\sqrt{(a+n)(a+n-1)}da,
\end{aligned}
\end{eqnarray}
from which we can get the E-Bayesian estimate of $\alpha$ is
\begin{eqnarray}
\hat{\alpha}_{EB2}=\frac{1}{c}\ln(\frac{T+c}{T})\int_{0}^{1}\sqrt{(a+n)(a+n-1)}da.
\end{eqnarray}

\subsection{E-Bayesian Estimation of $\alpha$ under EL function}
The E-Bayesian estimate of $\alpha$ under SEL function can be obtained from (2.5), (3.3), and (3.4) as follows:
\begin{eqnarray}
\begin{aligned}
\hat{\alpha}_{EB3}&=\iint_{Q}\hat{\alpha}_{B3}(a,b)\pi(a,b) \, da\,db
=\iint_{Q}\frac{a+n-1}{c(b+T)} \, da\,db\\&=\frac{1}{c}\int_{0}^{1}(a+n-1)da\int_{0}^{c}\frac{1}{b+T}db\\&=\frac{2n-1}{2c}\ln(\frac{T+c}{T}),
\end{aligned}
\end{eqnarray}
from which we can get the E-Bayesian estimate of $\alpha$ is
\begin{eqnarray}
\hat{\alpha}_{EB3}=\frac{2n-1}{2c}\ln(\frac{T+c}{T}),
\end{eqnarray}
where $T=\displaystyle\sum_{i=1}^n \ln(1+\frac{x_i}{\lambda})$.

\section{The E-MSE of E-Bayesian estimation }

We usually use MSE (mean square error) to measure the estimated error. The E-Bayesian estimation method proposed time is short, the research results obtained are also less,
especially the analytical expression of its estimated error has not been studied. \cite{Han2019b} in the case of the two hyperparameters, proposed the definition of
E-MSE. \cite{Han2019a}
in the case of the one hyperparameter, proposed the definition of E-MSE (expected mean
square error), too. From which we get the MSE of Bayesian estimation of $\hat{\alpha}_{BS}(a,b)$ as
\begin{eqnarray}
MSE[\hat{\alpha}_{B}(a,b)]=E\{[\alpha-\hat{\alpha}_{B} (a,b)]^2 |x\}.
\end{eqnarray}

So we can fix the E-Bayesian estimation of MSE called E-MSE as follows:
\begin{eqnarray}
E-MSE[\hat{\alpha}_{EB}(a,b)]=\iint_{Q}MSE[\hat{\alpha}_{B}(a,b)]\pi(a,b)\, da\,db,
\end{eqnarray}
which is the E-MSE (expected mean square error)) of E-Bayesian estimation $\hat{\alpha}_{EB}(a,b)$ where $Q$ is the
domain of a and b, $MSE[\hat{\alpha}_{B}(a,b)]$ is the MSE of Bayesian estimation of $\alpha$ with hyper
parameters $a$ and $b$, and $\pi(a, b)$ is the density function of $a$ and $b$ over $Q$. We indicates that $E-MSE[\hat{\alpha}_{EB}(a,b)]=E\{MES[\hat{\alpha}_{B}(a,b)]\}$. 

In this section, we will introduce the E-Bayesian estimation of $\alpha$ under different loss functions according to \cite{Han2020}.

\subsection{The E-MSE of $\hat{\alpha}_{EB}$ under SEL function }
The MSE of Bayesian Estimation $\hat{\alpha}(a,b)$ under SEL function can be obtained from (2.3), and (4.1) as follows:
\begin{eqnarray}
MSE[\hat{\alpha}_{B1}(a,b)]=E\{[\alpha-\hat{\alpha}_{B1} (a,b)]^2 |x\}.
\end{eqnarray}

According to proof procedure in section 3, the posterior distribution of $\alpha$
is $Gamma(n+a, T+b)$, so $Var(\alpha|x) = (n + a)/(T + b)^2$. The Bayesian estimation of $\alpha$ is $\hat{\alpha}_{B1}={(a+n)}/{(b+T)}$. So, the MSE of $\hat{\alpha}_{B1}$ is

\begin{eqnarray}
MSE[\hat{\alpha}_{B1}(a,b)]=E\{[\alpha-\hat{\alpha}_{B1} (a,b)]^2 |x\}=Var(\alpha|x) =\frac{n + a}{(T + b)^2}.
\end{eqnarray}

The prior density function $\pi(a, b)$ of $a$ and $b$ is given by (3.4),  then the E-MSE of E-Bayesian estimation $\hat{\alpha}_{EB1}$ is

\begin{eqnarray}
\begin{aligned}
E-MSE(\hat{\alpha}_{EB1})&=\iint_{Q}MSE[\hat{\alpha}_{B1}(a,b)]\pi(a,b) \, da\,db
\\&=\iint_{Q}\frac{n + a}{(T + b)^2}\frac{1}{c} \, da\,db\\&=\frac{1}{c}\int_{0}^{1}(a+n)da\int_{0}^{c}\frac{1}{(b+T)^2}db\\&
=\frac{2n+1}{2T(T+c)},
\end{aligned}
\end{eqnarray}
from which we can get the E-MSE of $\hat{\alpha}_{EB1}$ is
\begin{eqnarray}
E-MSE(\hat{\alpha}_{EB1})=\frac{2n+1}{2T(T+c)}.
\end{eqnarray}

\subsection{The E-MSE of $\hat{\alpha}_{EB}$ under KL function }
The MSE of Bayesian Estimation $\hat{\alpha}(a,b)$ under SEL function can be obtained from (2.4), and (4.1). When $\hat{\alpha}_{B2}={\sqrt{(a+n)(a+n-1)}}/{(b+T)}$,

\begin{eqnarray}
\begin{aligned}
MSE[\hat{\alpha}_{B2}(a,b)]&=E\{[\alpha-\hat{\alpha}_{B2}(a,b)]^2|x\}\\&=E(\alpha^2|x)-2\hat{\alpha}_{B2}(a,b)E(\alpha|x) + [\hat{\alpha}_{B2}(a,b)]^2\\&=\frac{2(n+a)[(n+a)-\sqrt{(a+n)(a+n-1)}]}{(b+T)^2}.
\end{aligned}
\end{eqnarray}

Then the E-MSE of E-Bayesian estimation $\hat{\alpha}_{EB2}$ is

\begin{eqnarray}
\begin{aligned}
E-MSE(\hat{\alpha}_{EB2})&=\iint_{Q}MSE[\hat{\alpha}_{B2}(a,b)]\pi(a,b) \, da\,db
\\&=\iint_{Q}\frac{2(n+a)[(n+a)-\sqrt{(a+n)(a+n-1)}]}{(b+T)^2}\frac{1}{c} \, da\,db\\&=\frac{1}{c}\int_{0}^{1}2(n+a)[(n+a)-\sqrt{(a+n)(a+n-1)}]da\int_{0}^{c}\frac{1}{(b+T)^2}db\\&
=\frac{2}{T(T+c)}\int_{0}^{1}(n+a)[(n+a)-\sqrt{(a+n)(a+n-1)}]da,
\end{aligned}
\end{eqnarray}\\
from which we can get the E-MSE of $\hat{\alpha}_{EB2}$ is
\begin{eqnarray}
E-MSE(\hat{\alpha}_{EB2})=\frac{2}{T(T+c)}\int_{0}^{1}(n+a)[(n+a)-\sqrt{(a+n)(a+n-1)}]da.
\end{eqnarray}

\subsection{The E-MSE of $\hat{\alpha}_{EB}$ under EL function }

The MSE of Bayesian Estimation $\hat{\alpha}(a,b)$ under SEL function can be obtained from (2.4), and (4.1). When $\hat{\alpha}_{EB3}={(a+n-1)}/{(b+T)}$,

\begin{eqnarray}
\begin{aligned}
MSE[\hat{\alpha}_{B3}(a,b)]&=E\{[\alpha-\hat{\alpha}_{B3}(a,b)]^2|x\}\\&=E(\alpha^2|x)-2\hat{\alpha}_{B3}(a,b)E(\alpha|x) + [\hat{\alpha}_{B3}(a,b)]^2\\&=\frac{a+n+1}{(b+T)^2}.
\end{aligned}
\end{eqnarray}

Then the E-MSE of E-Bayesian estimation $\hat{\alpha}_{EB3}$ is

\begin{eqnarray}
\begin{aligned}
E-MSE(\hat{\alpha}_{EB1})&=\iint_{Q}MSE[\hat{\alpha}_{B1}(a,b)]\pi(a,b) \, da\,db
\\&=\iint_{Q}\frac{n+a+1}{(T + b)^2}\frac{1}{c} \, da\,db\\&=\frac{1}{c}\int_{0}^{1}(a+n+1)da\int_{0}^{c}\frac{1}{(b+T)^2}db\\&
=\frac{2n+3}{2T(T+c)},
\end{aligned}
\end{eqnarray}
from which we can get the E-MSE of $\hat{\alpha}_{EB2}$ is
\begin{eqnarray}
E-MSE(\hat{\alpha}_{EB3})=\frac{2n+3}{2T(T+c)}.
\end{eqnarray}

\section{Monte Carlo simulation}
In this section, the Monte Carlo simulation is used to compare the E-Bayesian estimation and its E-MSE under different loss functions.
For known values of $\lambda=1,2,3$ and $\alpha=2.5$, a sample of size $n$ is then generated from
the Lomax distribution by using the inverse function method as follows: \begin{eqnarray}
X=\displaystyle[(1-U)^{-\frac{1}{\alpha}}-1]\lambda,
\end{eqnarray}
in which $U$ is generated from U(0,1). The codes of R are
used to generate datasets from the uniform and Lomax distributions.

The performance of all estimations are compared numerically by their $\hat{\alpha}_{EBi}$ and $E-MSE(\hat{\alpha}_{EBi})$
 $(i = 1, 2, 3)$ values. The simulation results with 10000 repetitions are given in table 1-3 $(c = 0.5, 1, 1.5)$.

Similarly, for known values of $\lambda=1,2,3$ and $\alpha=5.0$, a sample of size $n$ is then generated from
the Lomax distribution by using the inverse function method too. The performance of all estimates are
compared numerically by their $\hat{\alpha}_{EBi}$ and $E-MSE(\hat{\alpha}_{EBi})$ $(i = 1, 2, 3)$ values.
The simulation results with 10000 repetitions are given in table 4-6 $(c = 0.5, 1, 1.5)$.
Based on tabulated the values of $E-MSE(\hat{\alpha}_{EBi})$ $(i = 1, 2, 3)$ $(i = 1, 2, 3)$, the following
conclusions can be drawn from Tables 1–3 or Tables 4–6 (c = 0.5, 1, 1.5).

\begin{enumerate}[(1)]
\item Based on different loss functions, for the fixed $c (c = 0.5, 1, 1.5)$, it is observed that
by the increasing of $n$, the performances of all estimations improve in terms of $\hat{\alpha}_{EBi}$ and $E-MSE(\hat{\alpha}_{EBi})$ $(i = 1, 2, 3)$.
\item For the same values of $n(n = 20, 40, 60, 80, 100)$ and the different values of $c (c = 0.5, 1, 1.5)$, the values
of $\hat{\alpha}_{EBi}$ and  $E-MSE(\hat{\alpha}_{EBi})$ $(i = 1, 2, 3)$ are all robust.
\item For the same $n(n = 20, 40, 60, 80, 100)$ and $c (c = 0.5, 1, 1.5)$, the values of $\hat{\alpha}_{EBi}$ are close to each other, the values of $E-MSE(\hat{\alpha}_{EBi})$ $(i = 1, 2, 3)$ are close to
\end{enumerate}
\begin{eqnarray}
E-MSE(\hat{\alpha}_{EB1})< E-MSE(\hat{\alpha}_{EB2})< E-MSE(\hat{\alpha}_{EB3}).
\end{eqnarray}

It also suggests that, if E-MSE as the evaluation standard, then have conclusions as
follows: $\hat{\alpha}_{EB1}$ is more accurate than $\hat{\alpha}_{EB2}$, and $\hat{\alpha}_{EB2}$ is more accurate than $\hat{\alpha}_{EB3}$.

\begin{table}\small
	\centering
	\caption{Results of $\hat{\alpha}_{EBi}$ and $E-MSE(\hat{\alpha}_{EBi})$ $(i = 1, 2, 3; c = 0.5, 1, 1.5; \lambda=1; \alpha=2.5)$.}
	\begin{tabular}{c|c|ccc|ccc}
		\hline
	$n$&$c$	& $\hat{\alpha}_{EB1}$ & $\hat{\alpha}_{EB2}$ & $\hat{\alpha}_{EB3}$ & ${E-MSE}(\hat{\alpha}_{EB1})$ & $E-MSE(\hat{\alpha}_{EB2})$  & $E-MSE(\hat{\alpha}_{EB3})$  \\ \hline
		20&0.5 & 2.61178 & 2.54728 & 2.48438 & 0.35049 & 0.35488 & 0.36759 \\ 
		40& & 2.55534 & 2.52359 & 2.49224 & 0.16533 & 0.16637 & 0.16942 \\ 
		60& & 2.53703 & 2.51598 & 2.49510 & 0.10814 & 0.10859 & 0.10993 \\ 
		80& & 2.53287 & 2.51708 & 2.50140 & 0.08071 & 0.08096 & 0.08171 \\ 
		100& & 2.52099 & 2.50841 & 2.49590 & 0.06389 & 0.06405 & 0.06453 \\ \hline
		$n$&$c$	& $\hat{\alpha}_{EB1}$ & $\hat{\alpha}_{EB2}$ & $\hat{\alpha}_{EB3}$ & ${E-MSE}(\hat{\alpha}_{EB1})$ & $E-MSE(\hat{\alpha}_{EB2})$  & $E-MSE(\hat{\alpha}_{EB3})$ \\ \hline
		20&1 & 2.51231 & 2.45027 & 2.38976 & 0.32314 & 0.32718 & 0.33890 \\ 
		40& & 2.52068 & 2.48937 & 2.45844 & 0.16087 & 0.16188 & 0.16484 \\ 
		60& & 2.51319 & 2.49233 & 2.47165 & 0.10616 & 0.10660 & 0.10791 \\ 
		80& & 2.51219 & 2.49654 & 2.48098 & 0.07936 & 0.07961 & 0.08035 \\ 
		100& & 2.50503 & 2.49253 & 2.48010 & 0.06307 & 0.06322 & 0.06369 \\ \hline
		$n$&$c$	& $\hat{\alpha}_{EB1}$ & $\hat{\alpha}_{EB2}$ & $\hat{\alpha}_{EB3}$ & ${E-MSE}(\hat{\alpha}_{EB1})$ & $E-MSE(\hat{\alpha}_{EB2})$  & $E-MSE(\hat{\alpha}_{EB3})$ \\ \hline
		20&1.5 & 2.44681 & 2.38638 & 2.32745 & 0.30578 & 0.30960 & 0.32070 \\ 
		40& & 2.47814 & 2.44735 & 2.41695 & 0.15545 & 0.15642 & 0.15929 \\ 
		60& & 2.47948 & 2.45891 & 2.43850 & 0.10331 & 0.10374 & 0.10502 \\ 
		80& & 2.48935 & 2.47384 & 2.45843 & 0.07792 & 0.07816 & 0.07889 \\ 
		100& & 2.49199 & 2.47957 & 2.46720 & 0.06242 & 0.06257 & 0.06304 \\ \hline
	\end{tabular}
\end{table}

\begin{table}\small
	\centering
	\caption{Results of $\hat{\alpha}_{EBi}$ and $E-MSE(\hat{\alpha}_{EBi})$ $(i = 1, 2, 3; c = 0.5, 1, 1.5; \lambda=2; \alpha=2.5)$.}
	\begin{tabular}{c|c|ccc|ccc}
		\hline
	$n$&$c$	& $\hat{\alpha}_{EB1}$ & $\hat{\alpha}_{EB2}$ & $\hat{\alpha}_{EB3}$ & ${E-MSE}(\hat{\alpha}_{EB1})$ & $E-MSE(\hat{\alpha}_{EB2})$  & $E-MSE(\hat{\alpha}_{EB3})$ \\ \hline
		20 &0.5 &2.61821 & 2.55355 & 2.49049 & 0.35218 & 0.35659 & 0.36936\\ 
		40 & &2.54974 & 2.51807 & 2.48679 & 0.16470 & 0.16573 & 0.16877 \\ 
		60 & &2.53555 & 2.51451 & 2.49364 & 0.10803 & 0.10848 & 0.10982\\ 
		80 & &2.53315 & 2.51736 & 2.50168 & 0.08072 & 0.08097 & 0.08172 \\ 
		100 & &2.52438 & 2.51179 & 2.49926 & 0.06406 & 0.06422 & 0.06469  \\ \hline
		$n$&$c$	& $\hat{\alpha}_{EB1}$ & $\hat{\alpha}_{EB2}$ & $\hat{\alpha}_{EB3}$ & ${E-MSE}(\hat{\alpha}_{EB1})$ & $E-MSE(\hat{\alpha}_{EB2})$  & $E-MSE(\hat{\alpha}_{EB3})$  \\ \hline
		20 &1 &2.52610 & 2.46371 & 2.40287 & 0.32684 & 0.33093 & 0.34278 \\ 
		40 & &2.50950 & 2.47832 & 2.44754 & 0.15942 & 0.16042 & 0.16336\\ 
		60 & &2.51109 & 2.49025 & 2.46959 & 0.10602 & 0.10646 & 0.10777 \\ 
		80 & & 2.50994 & 2.49430 & 2.47876 & 0.07924 & 0.07949 & 0.08022 \\ 
		100 & &2.49940 & 2.48693 & 2.47453 & 0.06278 & 0.06294 & 0.06341 \\ \hline
		$n$&$c$	& $\hat{\alpha}_{EB1}$ & $\hat{\alpha}_{EB2}$ & $\hat{\alpha}_{EB3}$ & ${E-MSE}(\hat{\alpha}_{EB1})$ & $E-MSE(\hat{\alpha}_{EB2})$  & $E-MSE(\hat{\alpha}_{EB3})$ \\ \hline
		20 &1.5 &2.45259 & 2.39202 & 2.33295 & 0.30789 & 0.31174 & 0.32291 \\ 
		40 & &2.47615 & 2.44539 & 2.41501 & 0.15504 & 0.15601 & 0.15886  \\ 
		60 & &2.48266 & 2.46205 & 2.44162 & 0.10357 & 0.10400 & 0.10528 \\ 
		80 & & 2.48582 & 2.47033 & 2.45494 & 0.07769 & 0.07793 & 0.07865 \\ 
		100 & &2.49071 & 2.47829 & 2.46593 & 0.06234 & 0.06250 & 0.06296 \\ \hline
	\end{tabular}
\end{table}

\begin{table}\small
	\centering
	\caption{Results of $\hat{\alpha}_{EBi}$ and $E-MSE(\hat{\alpha}_{EBi})$ $(i = 1, 2, 3; c = 0.5, 1, 1.5; \lambda=3; \alpha=2.5)$.}
	\begin{tabular}{c|c|ccc|ccc}
		\hline
		$n$&$c$	& $\hat{\alpha}_{EB1}$ & $\hat{\alpha}_{EB2}$ & $\hat{\alpha}_{EB3}$ & ${E-MSE}(\hat{\alpha}_{EB1})$ & $E-MSE(\hat{\alpha}_{EB2})$  & $E-MSE(\hat{\alpha}_{EB3})$\\ \hline
		20 & 0.5 & 2.6076 & 2.54321 & 2.4804 & 0.3484 & 0.35275 & 0.36539 \\ 
		40 &  & 2.5563 & 2.52454 & 2.49318 & 0.16547 & 0.1665 & 0.16955 \\ 
		60 &  & 2.54242 & 2.52132 & 2.5004 & 0.10866 & 0.10912 & 0.11046 \\ 
		80 &  & 2.52736 & 2.51162 & 2.49597 & 0.08035 & 0.0806 & 0.08135 \\ 
		100 &  & 2.5214 & 2.50883 & 2.49631 & 0.06391 & 0.06407 & 0.06454 \\ \hline
		$n$&$c$	& $\hat{\alpha}_{EB1}$ & $\hat{\alpha}_{EB2}$ & $\hat{\alpha}_{EB3}$ & ${E-MSE}(\hat{\alpha}_{EB1})$ & $E-MSE(\hat{\alpha}_{EB2})$  & $E-MSE(\hat{\alpha}_{EB3})$\\ \hline
		20 &1  & 2.51429 & 2.4522 & 2.39165 & 0.32381 & 0.32786 & 0.33961 \\ 
		40 &  & 2.5168 & 2.48553 & 2.45466 & 0.16031 & 0.16131 & 0.16426 \\ 
		60 &  & 2.50769 & 2.48688 & 2.46624 & 0.10567 & 0.10611 & 0.10742 \\ 
		80 &  & 2.50809 & 2.49246 & 2.47693 & 0.07914 & 0.07938 & 0.08012 \\ 
		100 &  & 2.50936 & 2.49685 & 2.48439 & 0.06327 & 0.06343 & 0.0639 \\ \hline
		$n$&$c$	& $\hat{\alpha}_{EB1}$ & $\hat{\alpha}_{EB2}$ & $\hat{\alpha}_{EB3}$ & ${E-MSE}(\hat{\alpha}_{EB1})$ & $E-MSE(\hat{\alpha}_{EB2})$  & $E-MSE(\hat{\alpha}_{EB3})$\\ \hline
		20 & 1.5 & 2.44664 & 2.38622 & 2.32729 & 0.30654 & 0.31037 & 0.32149 \\ 
		40 &  & 2.48712 & 2.45623 & 2.42571 & 0.15654 & 0.15752 & 0.1604 \\ 
		60 &  & 2.48282 & 2.46222 & 2.44178 & 0.10359 & 0.10402 & 0.1053 \\ 
		80 &  & 2.48943 & 2.47392 & 2.45851 & 0.07794 & 0.07818 & 0.0789 \\ 
		100 &  & 2.49085 & 2.47842 & 2.46606 & 0.06235 & 0.0625 & 0.06297 \\ \hline
	\end{tabular}
\end{table}

\begin{table}\small
	\caption{Results of $\hat{\alpha}_{EBi}$ and $E-MSE(\hat{\alpha}_{EBi})$ $(i = 1, 2, 3; c = 0.5, 1, 1.5; \lambda=1; \alpha=5.0)$.}
	\begin{tabular}{c|c|ccc|ccc}
		\hline
		$n$&$c$	& $\hat{\alpha}_{EB1}$ & $\hat{\alpha}_{EB2}$ & $\hat{\alpha}_{EB3}$ & ${E-MSE}(\hat{\alpha}_{EB1})$ & $E-MSE(\hat{\alpha}_{EB2})$  & $E-MSE(\hat{\alpha}_{EB3})$\\ \hline
		20 & 0.5 & 5.03351 & 4.90920 & 4.78797 & 1.29946 & 1.31571 & 1.36285 \\ 
		40 &  & 5.03234 & 4.96982 & 4.90809 & 0.64105 & 0.64506 & 0.65688 \\ 
		60 &  & 5.01574 & 4.97412 & 4.93284 & 0.42262 & 0.42438 & 0.42961 \\ 
		80 &  & 5.01559 & 4.98434 & 4.95328 & 0.31635 & 0.31734 & 0.32028 \\ 
		100 &  & 5.01650 & 4.99148 & 4.96658 & 0.25294 & 0.25357 & 0.25546 \\ \hline
	$n$&$c$	& $\hat{\alpha}_{EB1}$ & $\hat{\alpha}_{EB2}$ & $\hat{\alpha}_{EB3}$ & ${E-MSE}(\hat{\alpha}_{EB1})$ & $E-MSE(\hat{\alpha}_{EB2})$  & $E-MSE(\hat{\alpha}_{EB3})$\\ \hline
		20 & 1 & 4.76401 & 4.64636 & 4.53162 & 1.16112 & 1.17563 & 1.21776 \\ 
		40 &  & 4.87868 & 4.81807 & 4.75822 & 0.60213 & 0.60590 & 0.61700 \\ 
		60 &  & 4.92782 & 4.88693 & 4.84637 & 0.40811 & 0.40981 & 0.41486 \\ 
		80 &  & 4.93723 & 4.90647 & 4.87590 & 0.30655 & 0.30751 & 0.31036 \\ 
		100 &  & 4.95220 & 4.92750 & 4.90292 & 0.24644 & 0.24706 & 0.24890 \\ \hline
		$n$&$c$	& $\hat{\alpha}_{EB1}$ & $\hat{\alpha}_{EB2}$ & $\hat{\alpha}_{EB3}$ & ${E-MSE}(\hat{\alpha}_{EB1})$ & $E-MSE(\hat{\alpha}_{EB2})$  & $E-MSE(\hat{\alpha}_{EB3})$\\ \hline
		20 & 1.5 & 4.51280 & 4.40136 & 4.29267 & 1.04374 & 1.05679 & 1.09465 \\ 
		40 &  & 4.74522 & 4.68627 & 4.62805 & 0.56960 & 0.57316 & 0.58367 \\ 
		60 &  & 4.82094 & 4.78093 & 4.74126 & 0.39061 & 0.39224 & 0.39706 \\ 
		80 &  & 4.87408 & 4.84371 & 4.81353 & 0.29877 & 0.29970 & 0.30248 \\ 
		100 &  & 4.88996 & 4.86557 & 4.84130 & 0.24028 & 0.24088 & 0.24267 \\ \hline
	\end{tabular}
\end{table}

\begin{table}\small
	\caption{Results of $\hat{\alpha}_{EBi}$ and $E-MSE(\hat{\alpha}_{EBi})$ $(i = 1, 2, 3; c = 0.5, 1, 1.5; \lambda=2; \alpha=5.0)$.}
	\begin{tabular}{c|c|ccc|ccc}
		\hline
		$n$&$c$	& $\hat{\alpha}_{EB1}$ & $\hat{\alpha}_{EB2}$ & $\hat{\alpha}_{EB3}$ & ${E-MSE}(\hat{\alpha}_{EB1})$ & $E-MSE(\hat{\alpha}_{EB2})$  & $E-MSE(\hat{\alpha}_{EB3})$\\ \hline
		20 & 0.5 & 5.05181 & 4.92705 & 4.80538 & 1.30710 & 1.32344 & 1.37086 \\ 
		40 &  & 5.03330 & 4.97078 & 4.90902 & 0.64126 & 0.64526 & 0.65709 \\ 
		60 &  & 5.01051 & 4.96892 & 4.92769 & 0.42194 & 0.42369 & 0.42891 \\ 
		80 &  & 5.01696 & 4.98570 & 4.95464 & 0.31666 & 0.31765 & 0.32060 \\ 
		100 &  & 5.01546 & 4.99045 & 4.96556 & 0.25285 & 0.25348 & 0.25536 \\ \hline
		$n$&$c$	& $\hat{\alpha}_{EB1}$ & $\hat{\alpha}_{EB2}$ & $\hat{\alpha}_{EB3}$ & ${E-MSE}(\hat{\alpha}_{EB1})$ & $E-MSE(\hat{\alpha}_{EB2})$  & $E-MSE(\hat{\alpha}_{EB3})$  \\ \hline
		20 & 1 & 4.76170 & 4.64411 & 4.52942 & 1.15905 & 1.17354 & 1.21559 \\ 
		40 &  & 4.87199 & 4.81147 & 4.75169 & 0.60066 & 0.60442 & 0.61549 \\ 
		60 &  & 4.91504 & 4.87425 & 4.83380 & 0.40603 & 0.40772 & 0.41274 \\ 
		80 &  & 4.93244 & 4.90171 & 4.87117 & 0.30599 & 0.30695 & 0.30979 \\ 
		100 &  & 4.95464 & 4.92993 & 4.90534 & 0.24666 & 0.24727 & 0.24911 \\ \hline
		$n$&$c$	& $\hat{\alpha}_{EB1}$ & $\hat{\alpha}_{EB2}$ & $\hat{\alpha}_{EB3}$ & ${E-MSE}(\hat{\alpha}_{EB1})$ & $E-MSE(\hat{\alpha}_{EB2})$  & $E-MSE(\hat{\alpha}_{EB3})$\\ \hline
		20 & 1.5 & 4.51377 & 4.40230 & 4.29359 & 1.04380 & 1.05685 & 1.09471 \\ 
		40 &  & 4.73634 & 4.67750 & 4.61939 & 0.56765 & 0.57120 & 0.58167 \\ 
		60 &  & 4.81990 & 4.77990 & 4.74024 & 0.39030 & 0.39193 & 0.39675 \\ 
		80 &  & 4.85921 & 4.82893 & 4.79884 & 0.29694 & 0.29787 & 0.30063 \\ 
		100 &  & 4.88936 & 4.86497 & 4.84071 & 0.24021 & 0.24081 & 0.24260 \\ \hline
	\end{tabular}
\end{table}

\begin{table}\small
	\caption{Results of $\hat{\alpha}_{EBi}$ and $E-MSE(\hat{\alpha}_{EBi})$ $(i = 1, 2, 3; c = 0.5, 1, 1.5; \lambda=2; \alpha=5.0)$.}
	\begin{tabular}{c|c|ccc|ccc}
		\hline
 	$n$&$c$	& $\hat{\alpha}_{EB1}$ & $\hat{\alpha}_{EB2}$ & $\hat{\alpha}_{EB3}$ & ${E-MSE}(\hat{\alpha}_{EB1})$ & $E-MSE(\hat{\alpha}_{EB2})$  & $E-MSE(\hat{\alpha}_{EB3})$ \\ \hline
		20 & 0.5 & 5.04171 & 4.91720 & 4.79577 & 1.30234 & 1.31863 & 1.36587 \\ 
		40 &  & 5.04674 & 4.98404 & 4.92213 & 0.64480 & 0.64883 & 0.66072 \\ 
		60 &  & 5.01089 & 4.96931 & 4.92807 & 0.42197 & 0.42373 & 0.42894 \\ 
		80 &  & 5.00780 & 4.97660 & 4.94560 & 0.31541 & 0.31639 & 0.31932 \\ 
		100 &  & 5.01213 & 4.98713 & 4.96226 & 0.25248 & 0.25311 & 0.25499 \\ \hline
		$n$&$c$	& $\hat{\alpha}_{EB1}$ & $\hat{\alpha}_{EB2}$ & $\hat{\alpha}_{EB3}$ & ${E-MSE}(\hat{\alpha}_{EB1})$ & $E-MSE(\hat{\alpha}_{EB2})$  & $E-MSE(\hat{\alpha}_{EB3})$ \\ \hline
		20 & 1 & 4.76227 & 4.64466 & 4.52996 & 1.16061 & 1.17512 & 1.21722 \\ 
		40 &  & 4.87384 & 4.81329 & 4.75349 & 0.60098 & 0.60474 & 0.61582 \\ 
		60 &  & 4.91167 & 4.87091 & 4.83049 & 0.40530 & 0.40699 & 0.41200 \\ 
		80 &  & 4.93223 & 4.90150 & 4.87096 & 0.30596 & 0.30691 & 0.30976 \\ 
		100 &  & 4.95563 & 4.93091 & 4.90632 & 0.24680 & 0.24742 & 0.24926 \\ \hline
		$n$&$c$	& $\hat{\alpha}_{EB1}$ & $\hat{\alpha}_{EB2}$ & $\hat{\alpha}_{EB3}$ & ${E-MSE}(\hat{\alpha}_{EB1})$ & $E-MSE(\hat{\alpha}_{EB2})$  & $E-MSE(\hat{\alpha}_{EB3})$  \\ \hline
		20 & 1.5 & 4.51528 & 4.40378 & 4.29503 & 1.04529 & 1.05836 & 1.09628 \\ 
		40 &  & 4.73029 & 4.67153 & 4.61350 & 0.56608 & 0.56962 & 0.58006 \\ 
		60 &  & 4.80995 & 4.77004 & 4.73045 & 0.38870 & 0.39032 & 0.39513 \\ 
		80 &  & 4.86516 & 4.83485 & 4.80472 & 0.29761 & 0.29854 & 0.30131 \\ 
		100 &  & 4.89147 & 4.86707 & 4.84280 & 0.24038 & 0.24099 & 0.24278 \\ \hline
	\end{tabular}
\end{table}

\section{Real data analysis} 
For the purpose of illustration, a real data set is analyzed in this section. Numerical examples present the applicability of E-Bayesian estimation for Lomax distribution
based on three loss functions. Therefore, we investigate the performance of these methods when applied to
real data by estimating the parameter $\alpha$ of Lomax distribution.

\subsection{An Illustrative Example}
A numerical example is provided to verify the applicability of the above proposed methods. We investigate the performance of these methods when applied to real data sets by estimating the parameter $\alpha$ and the reliability function $R(t)$ of Lomax distribution. These data sets were provided by \cite{Bennett2000} and represent the minority electron mobility (the electron mobility describes how fast an electron can move through a metal or semiconductor under the action of an electric field) for P-type
(called P-type because holes carry positive (P) charge) $Ga_{1-x}Al_xAs$ (semiconductor of
Gallium Arsenite and Ammonium Arsenite) with seven different values of mole fraction. These data were obtained in the Semiconductor Electronics Division of the
National Institute of Standards and Technology Electronics and Electrical Engineering
Laboratory. Data set related to the mole fractions 0.25are used in this article. The real data set contain 21 observations are shown as below:
\\ \textbf{Real Data Set (belongs to mole fraction 0.25)}: 3.051, 2.779, 2.604, 2.371, 2.214, 2.045, 1.715, 1.525, 1.296, 1.154, 1.016, 0.7948, 0.7007, 0.6292, 0.6175, 0.6449, 0.8881, 1.115, 1.397, 1.506, 1.528.

To analyze these data sets quantitively, the Kolmogorov–Smirnov (K-S) test is applied. Kolmogorov-Smirnov test is based on cumulative distribution function in order to test whether two empirical distributions are different or whether one empirical distribution is different from another ideal distribution. K-S test checks if two independent distributions are similar or different, by generating cumulative probability plots for two distributions and finding the distance along the $y$-axis for given $x$ values between the two curves. From all the distances calculated for each $x$ value, we get the maximum distance. Then the maximum distance or maximum difference is plugged into the K-S probability function to calculate the probability value (p-value).  The lower the probability value is, the less likely the two distributions are similar.  Conversely, the higher or more close to 1 the value is, the more similar the two distributions are. For more details about K-S test, we can refer to \cite{Lopes2014}.
To test whether the Lomax distribution can be used to analyze the
datasets, Kolmogorov– Smirnov(K-S) tests are conducted. Since the K-S distances and the associated
p-values is (0.2857, 0.365) . For \textbf{Real Data Set}, we cannot
reject the null hypothesis that the Lomax distribution provides a reasonable fit.

\begin{table}\small
	\centering
	\caption{Results of $\hat{\alpha}_{EBi}$ and $E-MSE(\hat{\alpha}_{EBi})$ for Real Data Set.}
	\begin{tabular}{c|ccc|ccc}
		\hline
		$c$	& $\hat{\alpha}_{EB1}$ & $\hat{\alpha}_{EB2}$ & $\hat{\alpha}_{EB3}$ & ${E-MSE}(\hat{\alpha}_{EB1})$ & $E-MSE(\hat{\alpha}_{EB2})$  & $E-MSE(\hat{\alpha}_{EB3})$\\ \hline
		0.25 & 2.56133 & 2.50106 & 2.44220 & 0.30516 & 0.30879 & 0.31935 \\ 
		0.50 & 2.52429 & 2.46489 & 2.40688 & 0.29646 & 0.29999 & 0.31025 \\ 
		0.75 & 2.48864 & 2.43007 & 2.37289 & 0.28824 & 0.29167 & 0.30165 \\ 
		1.00 & 2.45429 & 2.39653 & 2.34013 & 0.28047 & 0.28381 & 0.29351 \\ 
		1.25 & 2.42116 & 2.36418 & 2.30855 & 0.27310 & 0.27635 & 0.28581 \\ \hline
	\end{tabular}
\end{table}

\begin{figure}[h]
	\centering
	\includegraphics[width=10cm,height=10cm]{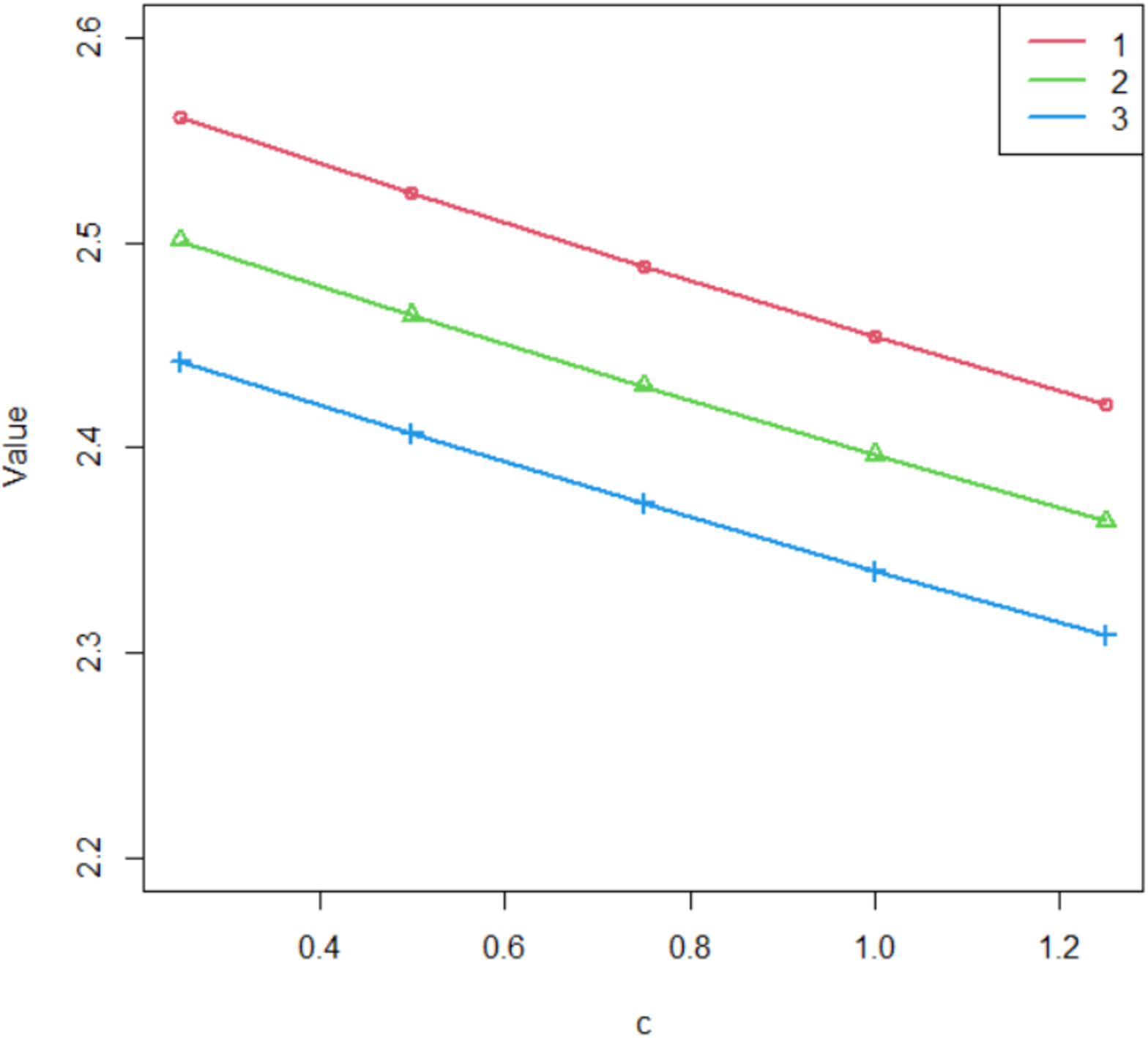}
	\caption{Relationship between $c$ and $\hat{\alpha}_{EBi}$$(i = 1, 2, 3)$}
\end{figure}

\begin{figure}[h]
	\centering
	\includegraphics[width=10cm,height=10cm]{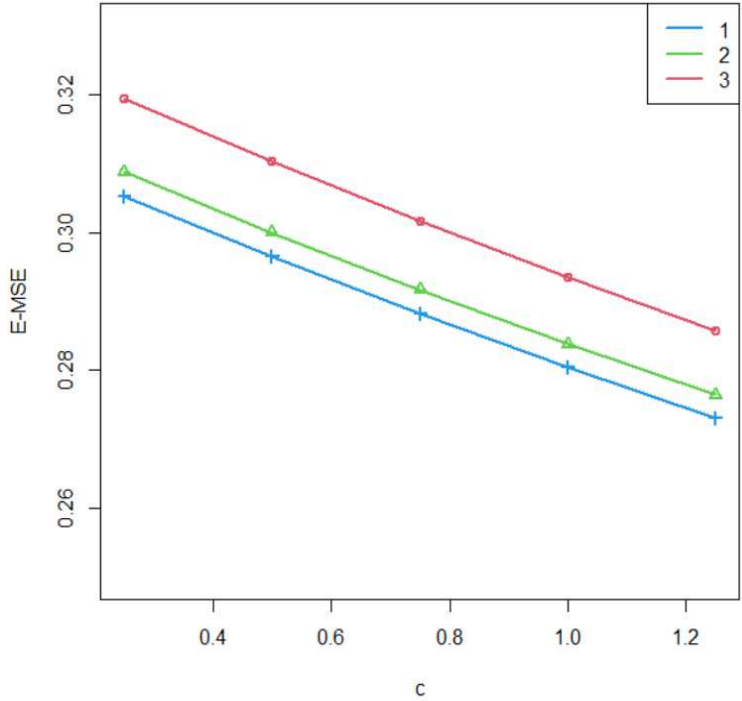}
	\caption{Relationship between $c$ and $E-MSE(\hat{\alpha}_{EBi})$$(i = 1, 2, 3)$}
\end{figure}

According to section 3, 4 and Table 7, when $\lambda=3$ we can get $\hat{\alpha}_{EBi}$ and $E-MSE(\hat{\alpha}_{EBi})$
$(i = 1, 2, 3)$ for Real Data Set 1.

From Table 7, we find that for the same $c (0.25, 0.50, 0.75, 1.00, 1.25)$, the values of $\hat{\alpha}_{EBi}$ are close to each other, and the values of $E-MSE(\hat{\alpha}_{EBi})$
$(i = 1, 2, 3)$ have the following sequential relationship:
 \begin{eqnarray}
E-MSE(\hat{\alpha}_{EB1})< E-MSE(\hat{\alpha}_{EB2})< E-MSE(\hat{\alpha}_{EB3}),
\end{eqnarray}
 moreover, for the different $c (0.25, 0.50, 0.75, 1.00, 1.25)$, the values of
$\hat{\alpha}_{EBi}$ and $E-MSE(\hat{\alpha}_{EBi})$
$(i = 1, 2, 3)$ are all robust.

From Figure 4, we also find that for the same $c (0.25, 0.50, 0.75, 1.00, 1.25)$, values of
 $E-MSE(\hat{\alpha}_{EBi})$
$(i = 1, 2, 3)$ have the following sequential relationship:
\begin{eqnarray}
E-MSE(\hat{\alpha}_{EB1})< E-MSE(\hat{\alpha}_{EB2})< E-MSE(\hat{\alpha}_{EB3}).
\end{eqnarray}
If $E-MSE$ is the evaluation standard, we have conclusions as follows: $E-MSE(\hat{\alpha}_{EB1})$ is more accurate than
$E-MSE(\hat{\alpha}_{EB2})$, $E-MSE(\hat{\alpha}_{EB2})$ is more accurate than $E-MSE(\hat{\alpha}_{EB2})$.

\subsection{Maximum likelihood estimation}

According to (1.12) the MLE of $\alpha$ is $\hat{\alpha}_{ML}={n}/{T}$. So the MLE of parameter $\alpha$ for the real data set is given by

\begin{eqnarray}
\hat{\alpha}_{ML}=\frac{n}{\displaystyle\sum_{i=1}^n \ln(1+\frac{x_i}{\lambda})}.
\end{eqnarray}

When $n=21$, $\lambda=3$,
\begin{eqnarray}
\hat{\alpha}_{ML}=2.539392.
\end{eqnarray}
as $c>0.25$,
\begin{eqnarray}
\hat{\alpha}_{ML}>\hat{\alpha}_{EB1}>\hat{\alpha}_{EB2}>\hat{\alpha}_{EB3}.
\end{eqnarray}
and as $c>0.25$,
\begin{eqnarray}
\hat{\alpha}_{EB1}>\hat{\alpha}_{EB2}>\hat{\alpha}_{EB2}>\hat{\alpha}_{EB3}.
\end{eqnarray}

We find that the values of $\hat{\alpha}_{EBi}$$(i = 1, 2, 3)$ and $\hat{\alpha}_{ML}$ are close to each other.

\section{Conclusion}

In this article, we apply Bayesian and E-Bayesian methods for estimating the parameter
$\alpha$ of Lomax distribution under r different loss functions (including squared error loss, entropy loss, and K-loss). Then E-MSE is introduced based on E-Bayesian estimation in order to measure the estimated error, and the formulas of E-Bayesian estimation and E-MSE are given. E-Bayesian estimations of $\hat{\alpha}_{EBi}$$(i = 1, 2, 3)$ and $E-MSE(\hat{\alpha}_{EBi})$
$(i = 1, 2, 3)$ under different loss functions are computed by calculus directly.

In this paper the maximum likelihood estimation $\hat{\alpha}_{ML}$ is also given, which is compared with the E-Bayesian estimation $\hat{\alpha}_{EBi}$$(i = 1, 2, 3)$ .

Reviewing the simulation example and application example, we find that the values of
$E-MSE(\hat{\alpha}_{EBi})$
$(i = 1, 2, 3)$ have the following sequential relationship:
\begin{eqnarray}
E-MSE(\hat{\alpha}_{EB1})< E-MSE(\hat{\alpha}_{EB2})< E-MSE(\hat{\alpha}_{EB3}).
\end{eqnarray}

It suggests that, if E-MSE is the evaluation standard, then we have the following
conclusions: $E-MSE(\hat{\alpha}_{EB1})$ is more accurate than
$E-MSE(\hat{\alpha}_{EB2})$, $E-MSE(\hat{\alpha}_{EB2})$ is more accurate than $E-MSE(\hat{\alpha}_{EB2})$.
When considering the E-Bayesian estimation problem under different loss functions, this paper proposes E-MSE as the evaluation standard.


\bibliographystyle{agsm}
\bibliography{jiaoda11}

\end{document}